# Tm-doped Crystals for mid-IR Optical Cryocoolers and Radiation Balanced Lasers

SAEID ROSTAMI[1], ALEXANDER R. ALBRECHT[1], AZZURRA VOLPI[1], MARKUS P. HEHLEN[1,2], MAURO TONELLI[3], AND MANSOOR SHEIK-BAHAE[1,*]

[1]Department of Physics and Astronomy, University of New Mexico, 1919 Lomas Blvd NE, MSC 07-4220, Albuquerque, NM 87131, USA
[2]Los Alamos National Laboratory, P.O. Box 1663, Los Alamos, NM 87545, USA
[3]NEST-CNR, Dipartimento di Fisica, Universita` di Pisa, Largo B. Pontecorvo 3, 56127 Pisa, Italy
*Corresponding author: msb@unm.edu

**We report the complete characterization of various cooling grade Tm-doped crystals including the first demonstration of optical refrigeration in Tm:YLF crystals. Room temperature laser cooling efficiencies of 1% and 2% (mol) Tm:YLF, and 1% Tm:BYF crystals at different excitation polarizations are measured and their external quantum efficiency and background absorption are extracted. By performing detailed low-temperature spectroscopic analysis of the samples, global minimum achievable temperatures of 160 K to 110 K are estimated. The potential of Tm-doped crystals to realize mid-IR optical cryocoolers and radiation balanced lasers (RBLs) in the eye-safe region of the spectrum is discussed, and a promising 2-tone RBL in a tandem structure of Tm:YLF and Ho:YLF crystals is proposed.**

Optical refrigeration was first demonstrated in Yb-doped ZBLANP glass in 1995 [1]. Since then, considerable advances have been achieved in solid-state optical cooling, approaching liquid Nitrogen temperature [2–4]. Recently, a Yb:YLF crystal was cooled from room temperature to below 90 K [4], and later an all-solid-state, vibration-free, cryogenic refrigerator device was realized based on Yb:YLF crystal that cooled a HgCdTe infrared (IR) sensor to 135 K [5]. Solid-state optical refrigeration relies on anti-Stokes fluorescence in which low entropy laser light with photon energy $h\nu$ less than mean fluorescence energy $h\nu_f$ is absorbed followed by efficient spontaneous emission that carries heat and entropy away from the system (Fig. 1a). A temperature ($T$) dependent cooling efficiency $\eta_c$ describing the ratio of the heat-lift to the absorbed laser power is expressed as [2]:

$$\eta_c(\lambda, T) = p(\lambda, T)\frac{\lambda}{\lambda_f(T)} - 1 \quad (1)$$

where $\lambda = c/\nu$ is the laser excitation wavelength, $\lambda_f$ is the mean fluorescence wavelength, and $p \lesssim 1$ describes the probability that an absorbed pump photon is converted to a fluorescence photon that exits the system [6]. It is straightforward to show that $p = \eta_{ext}\eta_{abs}$ with $\eta_{ext}$ describing the external quantum efficiency (EQE) and $\eta_{abs}$ the resonant absorption efficiency [2]. $\eta_{ext}$ describes the fraction of the excited ions that lead to radiative decay exiting the crystal. It is typically less than unity due to the omnipresence of nonradiative decay rate ($W_{nr}$) so that $\eta_{ext} = 1/(1 + W_{nr}/\eta_e W_r)$ [2]. Here, $W_r$ is the spontaneous emission rate, and $\eta_e$ is fluorescence escape efficiency that is typically less than unity because of fluorescence trapping and re-absorption due to total-internal reflections. Particularly for solid-state optical refrigeration in the mid-IR, choosing a host crystal with a low-phonon energy such as in fluorides is key to realizing high $\eta_{ext}$ (>98%) that is required for achieving net cooling ($\eta_c$>0). The quantities determining $\eta_{ext}$ are only weakly temperature dependent, and therefore $\eta_{ext}$ is taken to be a constant for a given material. The temperature dependence of $\eta_c$ is primarily due to $\eta_{abs}$ and $\lambda_f$ as a result of the Boltzmann distribution of thermal populations in the ground- and excited-state multiplets, respectively. More specifically, $\eta_{abs} = 1/(1 + \alpha_b/\alpha_r(\lambda, T))$ where $\alpha_r(\lambda, T)$ is the resonant absorption coefficient of the rare-earth ion, and $\alpha_b$ is the coefficient of parasitic (background) absorption due to presence of undesired impurities (such as transition metals) and other contaminants. $\alpha_b$ has been taken to be broadband (wavelength independent) within the spectral region of interest as well as temperature independent. Recent experiments in Yb:YLF crystals however have revealed that the generality of this assumption needs to be revisited [7]. The implications of this temperature dependency in $\alpha_b$ for mid-IR optical refrigeration will be addressed later in this letter.

Investigation of optical refrigeration in mid-IR has also been touted as a mean to enhance the overall cooling efficiency [8,9]. This follows the assumption that the maximum heat lift per absorbed photon is of the order of $k_B T$ where $k_B$ is the Boltzmann constant and $T$ is the lattice temperature. Such a prospect was analyzed in details more recently for Ho:YLF crystals [6]. In this letter, we report the first demonstration of optical refrigeration in Tm:YLF crystals and present complete low-temperature analysis of Tm-doped YLF and BYF crystals. We discuss the potential of these materials for mid-IR optical cryocoolers as well as their intriguing application for mid-IR RBLs by proposing 2-tone (2T)-RBL as a promising and novel candidate for high power operation of such *athermal* lasers. Anti-Stokes fluorescence cooling in Tm-doped crystals involves the ground state multiplet ($^3H_6$) and the first excited state multiplet ($^3F_4$) of $Tm^{3+}$ as shown in Fig. 1a. In a crystal field of sufficiently low symmetry, these multiplets can split into a maximum of 13 and 9 crystal-field levels, respectively. In higher symmetry such as in YLF, some levels remain degenerate, and only 10 and 7 crystal-field levels are observable, respectively [10].

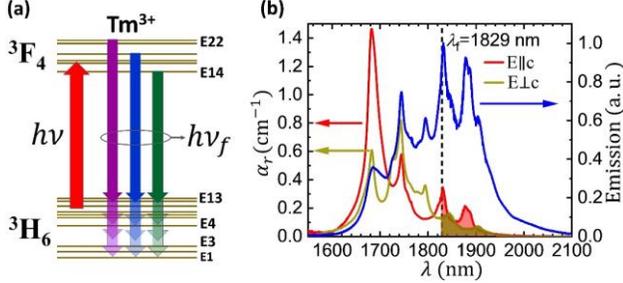

Fig. 1. (a) Anti-Stokes fluorescence cooling process in Tm:YLF crystal, (b) absorption spectra for E∥c and E⊥c polarizations (left axis) and emission spectrum of 2% Tm:YLF crystal (right axis) at *T*=300 K.

Laser cooling is observed if the laser excitation wavelength is tuned to within the so called 'cooling tail' (the shaded regions in Fig. 1b), *i.e.* if the crystal is excited with a laser of wavelength $\lambda>\lambda_f$. Fig. 1b shows the emission spectrum of a 2% Tm:YLF crystal, measured using a mid-IR optical spectrum analyzer (OSA), and the absorption spectra (for E∥c and E⊥c polarizations, where c represents the optical axis of the crystal), measured using a Fourier transform infrared (FTIR) spectrometer. Averaging over the two different polarized emission spectra, as required for uniaxial crystals [6], we obtain $\lambda_f$=1829 nm for 2% Tm:YLF at *T*=300 K. For Tm:BYF, a biaxial crystal, this averaging must be performed over six unique polarized emission spectra emanating from crystallographically oriented facets of the crystal [11]. To determine if a sample is of "cooling grade" and to quantify its external quantum efficiency ($\eta_{ext}$) and background absorption ($\alpha_b$), a coherent source tunable in the spectral vicinity of $\lambda_f$ is needed. For this purpose, we designed and constructed a temperature-tuned (1720-2800 nm) singly-resonant continuous-wave optical parametric oscillator (CW-OPO) based on 5 mol% MgO-doped periodically-polled lithium niobate (PPLN) crystal, achieving an output power of 1-2 W and a narrow linewidth (<1 nm) [6].

High-purity rare-earth doped fluoride single-crystals suitable for laser cooling are not readily available commercially. The crystals investigated in this work were grown using the Czochralski method from 5N (99.999%) purity (with respect to the total rare-earth concentration; and having low concentration of transition metals) binary fluorides. A 4.8(a)×4.8(a)×5.0(c) $mm^3$ sample was obtained from a 2% Tm:YLF crystal grown by AC Materials (Tarpon Springs, FL) using a 1 mm/h pulling and 7 rpm rotation rates under Nitrogen atmosphere. 2.75(a)×3.05(a)×3.2(c) $mm^3$ 1% Tm:YLF and 3.64(a)×5.25(b)×5.01(c) $mm^3$ 1% Tm:BYF samples were obtained from crystals grown at Pisa (Italy) using 1 and 0.5 mm/h pulling rates, respectively, and a 5 rpm rotation rate under Argon atmosphere. The biaxial 1% Tm:BYF sample was cut along the crystal axes (a, b, c) rather than the optical axes (X, Y, Z) in which b∥Y and ∠c, Z~22°, ∠a, c~99.76° [12].

The cooling efficiency $\eta_c$ of the 1% and 2% Tm:YLF crystals and the 1% Tm:BYF crystal were measured at room temperature using the method of Laser-Induced Thermal Modulation Spectroscopy (LITMoS) [13], as shown schematically in Fig. 2a.

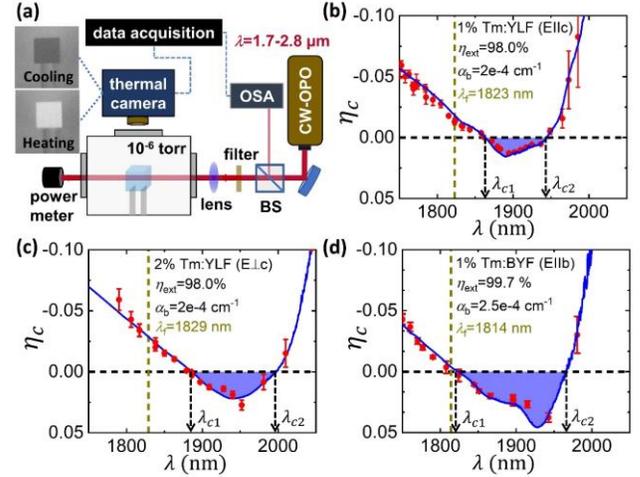

Fig. 2. (a) Schematic of LITMoS test setup, and the results of the test for (b) 1% Tm:YLF (E∥c), (c) 2% Tm:YLF (E⊥c), and (d) 1% Tm:BYF (E∥b).

In this technique, the cooling efficiency (at room temperature) $\eta_c(\lambda) = \Delta T/KP_{abs}$ is evaluated by measuring the laser-induced temperature change $\Delta T$ normalized to the absorbed power $P_{abs}$ as the laser (OPO) wavelength is tuned from below to above $\lambda_f$. $K$ is a scaling constant that varies inversely with the thermal load on the sample [6]. To minimize the thermal load on the sample, it is supported by two low thermal conductivity holders (microscope slide coverslips), and the experiment is performed in vacuum ($10^{-6}$ torr). $\Delta T$ is extracted from images captured by a thermal camera (L3-Communications Corporation, TX, USA), and $P_{abs}$ is evaluated from the incident and transmitted OPO powers as well as the known absorption spectra of the samples under study. Once $\eta_c(\lambda, T)$ is evaluated at *T*=300 K, the external quantum efficiency $\eta_{ext}$ and the background absorption $\alpha_b$ are obtained from a least-squares fit to Eq. 1.

**Table 1. External quantum efficiency $\eta_{ext}$, background absorption $\alpha_b$, global minimum achievable temperature (MAT), and optimum laser cooling wavelength $\lambda_{opt}$ for Tm-doped crystals at different excitation polarizations.**

| Sample | Pol. | $\eta_{ext}$ (%) | $\alpha_b$ ($10^{-4}$ cm$^{-1}$) | Global MAT (K) | $\lambda_{opt}$ (nm) |
|---|---|---|---|---|---|
| 1% Tm:YLF | E∥c | 98.0±0.2 | 2±1 | 190±10 | 1888 |
| 2% Tm:YLF | E∥c | 98.2±0.3 | 3±1 | 180±10 | 1889 |
| 2% Tm:YLF | E⊥c | 98.0±0.3 | 2±1 | 160±10 | 1910 |
| 1%Tm:BYF | E∥b | 99.7±0.2 | 2.5±0.5 | 160±10 | 1859 |
| 1%Tm:BYF | E∥c | 99.5±0.2 | 1.0±0.5 | 160±10 | 1857 |

Figures 2b-d show the LITMoS test results for the 1% Tm:YLF (E∥c), 2% Tm:YLF (E⊥c), and 1% Tm:BYF (E∥b) samples, respectively. As seen, each cooling-grade sample exhibits a spectral cooling window (the shaded regions in Fig. 2b-d) between two zero-crossing wavelengths $\lambda_{c1}$ and $\lambda_{c2}$ ($>\lambda_{c1}$). To a fair approximation $\eta_{ext} \approx \lambda_f/\lambda_{c1}$, and $\alpha_b$ depends on $\lambda_{c2}$, but its value is more accurately extracted by the fitting procedure [13]. Table 1 summarizes the LITMoS test results for Tm:YLF and Tm:BYF crystals at different excitation polarizations.

A figure of merit for a material intended for optical cryocooler applications is its Minimum Achievable Temperature (MAT), which describes the lowest temperature at which $\eta_c$ becomes zero [6,13]. The MAT is estimated by analyzing the temperature dependence of the constituents of $\eta_c$, i.e. $\lambda_f(T)$ and $\alpha_r(\lambda, T)$, and assuming that $\eta_{ext}$ and $\alpha_b$ are temperature independent [6], as stated earlier. $\lambda_f(T)$ is obtained from a weighted average of distinguishable polarized emission spectra $S(\lambda)$ of the crystal at each temperature. Figure 3a shows $\lambda_f(T)$ for 1% and 2% Tm:YLF and 1%Tm:BYF crystals from 300 K to 80 K. Note that $\lambda_f(T)$ for 2% Tm:YLF is ~9 nm greater than for 1% Tm:YLF due to the stronger reabsorption effect at higher doping concentrations. Both crystals show a ~3.4% red-shift in $\lambda_f(T)$ from 300 K to 80 K. The red-shift for the 1% Tm:BYF crystals is ~3.6% due to the greater crystal-field splitting of the multiplets in BYF compared to YLF [10]. The temperature-dependent resonant absorption coefficient spectra $\alpha_r(\lambda, T)$ are obtained from fluorescence spectra and applying the reciprocity theorem [14]. Their absolute magnitudes are calibrated to a measured absorption coefficient (e.g. at λ=1830 nm for Tm:YLF) using an FTIR spectrometer. The absorption spectra obtained from reciprocity exhibit less noise in the long wavelength tail (λ>1850 nm) compared to absorption spectra obtained from a direct absorption measurement. This is particularly relevant for laser cooling and RBL applications. Figures 3b-d show measured $\alpha_r(\lambda, T)$ spectra for 2% Tm:YLF (E∥c), 2% Tm:YLF (E⊥c) and 1% Tm:BYF (E∥b) crystals from 300 K to 80 K in 20 K increments.

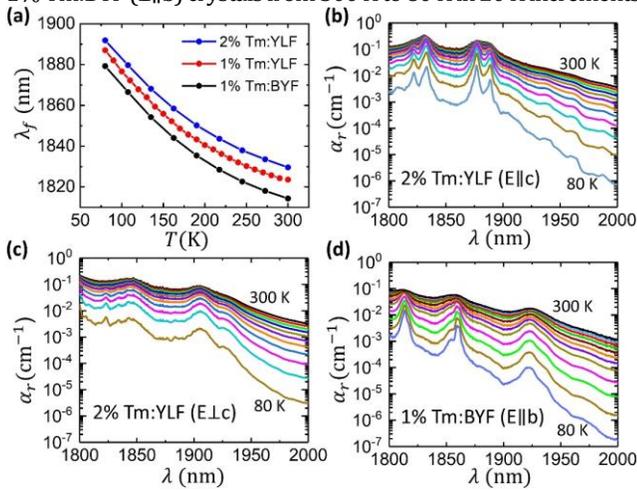

Fig. 3. (a) Temperature dependent mean fluorescence wavelength $\lambda_f(T)$ of 1% and 2% Tm:YLF and 1% Tm:BYF crystals, and temperature-dependent resonant absorption spectra $\alpha_r(\lambda, T)$ for (b) 2% Tm:YLF (E∥c), (c) 2% Tm:YLF (E⊥c), and (d) 1% Tm:BYF (E∥b) crystals from 300 K to 80 K in 20 K increments.

With $\eta_{ext}$, $\alpha_b$, $\lambda_f(T)$, and $\alpha_r(\lambda, T)$ obtained from experiment, Eq. 1 can be used to plot $\eta_c(\lambda, T)$ for each sample [6], as shown in Fig. 4. The white demarcation between cooling (blue) and heating (red) regimes shows the MAT for each wavelength, and the global MATs and their corresponding wavelengths marked by dashed lines. The global MATs for all characterized crystals at different excitation polarizations are reported in Table 1 (including those that are not shown graphically in Figs. 2-4). The 1% and 2% Tm:YLF crystals exhibit almost identical $\eta_{ext}$~98.0% and $\alpha_b$~2×10$^{-4}$ cm$^{-1}$, and the same shift in $\lambda_f(T)$; however, the 2% doped sample for E⊥c excitation polarization has a lower MAT of 160±10 K due to a greater $\alpha_r$ at higher doping concentration and the presence of a pronounced absorption peak around ~1910 nm as shown in Fig. 3(c). The 1% Tm:BYF sample has an extremely high external quantum efficiency ($\eta_{ext}$>99%), which we believe to be due to the lower phonon energy (350 cm$^{-1}$ in BYF compared to 450 cm$^{-1}$ in YLF) that further suppresses multi-phonon relaxation [15]. As a result, Tm:BYF offers a higher maximum cooling efficiency (~5% vs. ~2%) compared to Tm:YLF (Fig. 2b-d). Despite the smaller resonant absorption in Tm:BYF, this leads to a MAT (~160 K) comparable to 2% Tm:YLF (E⊥c). The estimated MATs could be improved by increasing the dopant concentration or by reducing the background absorption coefficient $\alpha_b$. The Tm$^{3+}$ dopant density could be increased to 3-4% without affecting $\eta_{ext}$, since quenching effects only become active at such doping concentrations [16]. The background absorption coefficient $\alpha_b$ could be lowered by further improvements to material purity and crystal growth. In addition, recent temperature-dependent investigations of the background absorption in Yb:YLF crystals show that $\alpha_b$ can decrease with a decreasing temperature of the crystal [7]. If such a behavior was generalized to Tm-doped crystals, the actual global MATs might be even lower than the estimated values shown in Table 1. With this assumption and (or) assuming improved material purity, the estimated global MATs for $\alpha_b$=1×10$^{-5}$ cm$^{-1}$ for 2% Tm:YLF (E⊥c) and 1% Tm:BYF (E∥b) and (E∥c) are ~110 K at a wavelength of ~1929 nm and ~1926 nm, respectively. This shows that Tm-doped crystals have the potential to enable mid-IR optical cryocoolers. Tm:BYF crystals offer the added benefit of not requiring a polarized excitation because the MATs and their corresponding wavelengths are essentially the same for both E∥b and E∥c polarizations.

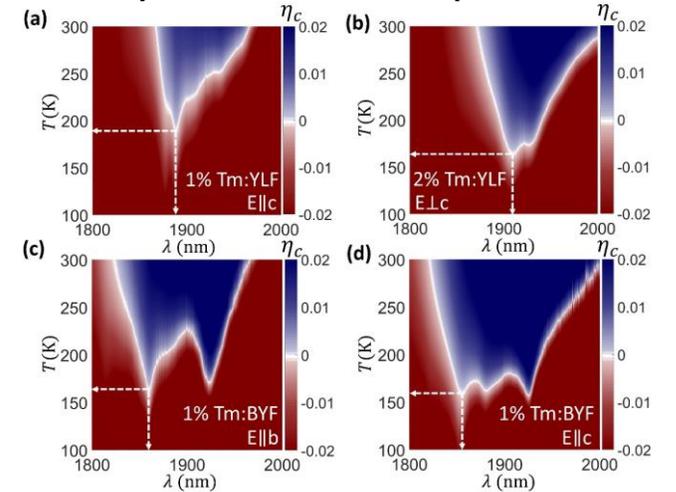

Fig. 4. Minimum Achievable Temperature (MAT) curves for (a) 1% Tm:YLF (E∥c), (b) 2% Tm:YLF (E⊥c), (c) 1% Tm:BYF (E∥b), and (d) 1% Tm:BYF (E∥c) crystals. The global MATs are marked (with white arrows) here and the corresponding values are reported in Table 1.

It is worth noting that mid-IR optical cryocoolers could potentially be used to cool loads that are transparent to mid-IR radiation (such as silicon based reference cavities [17,18]) without the need for an elaborate thermal-link as was needed in Ref. [5]. Also, $\lambda_{opt}$ of Tm-doped crystals coincide with readily available high-power Tm (fiber) lasers [19], which promises enhanced wall-plug efficiency compared to Yb-based laser cooling systems.

Cooling-grade crystals can also be used as gain media for *athermal* or Radiation Balanced Lasers (RBLs), where up-conversion fluorescence cooling balances the heat generated in a solid-state laser operating at a laser wavelength $\lambda_L$ [20,21]. As proposed by Bowman [20], the pump laser at $\lambda_P > \lambda_{c1}$ creates inversion and gain at $\lambda_L > \lambda_P$ while cooling the crystal. Given the gain/cooling material properties, one can find pump and laser intensities that satisfy the RBL condition for which the heat generated by the laser quantum defect is balanced by the fluorescence cooling. To date, RBL has been demonstrated in rod and disk geometries in the near-IR using Yb:YAG crystal [22,23]. Here we discuss the potential of Tm-doped crystals to realize mid-IR RBLs. Our experimental results show that the Tm:BYF crystal with $\eta_{ext}$ >99% is a highly promising candidate for realizing a mid-IR RBL. Fig. 5a shows a proposed pumping scheme for such a mid-IR RBL using a 1% Tm:BYF crystal with $\lambda_{c1}$=1817 nm, $\lambda_P$=1857 nm, and $\lambda_L$~1920 nm. The presence of an 1857 nm absorption resonance (corresponding to the $E_{11} \rightarrow E_{16}$ crystal-field transition) in Tm:BYF offers a suitable pumping wavelength with a large detuning of ~40 nm from $\lambda_{c1}$, making this crystal especially attractive for mid-IR RBLs [24]. In particular, if Tm:BYF crystals with high $Tm^{3+}$ doping concentrations (2-3%) are synthesized with $\eta_{ext}$>99%, high power RBLs with minimal thermal instabilities could be realized.

Another intriguing possibility in exploiting optical refrigeration in Tm-doped crystals is the idea of a 2-tone (2T)-RBL in conjunction with a Ho-doped system. 2T-RBLs, in general, refer to systems where the cooling and the lasing correspond to two separate transitions either in the same or different ions [25]. A Tm/Ho system is an ideal platform to realize a 2T-RBL in mid-IR. In this configuration, optical refrigeration in $Tm^{3+}$ balances the heat generated by the laser quantum defect in $Ho^{3+}$. Such a 2T-RBL could potentially be realized either in a cooling-grade Tm/Ho co-doped crystal or in a tandem (or distributed) geometry where Tm-doped cooling crystals are physically bonded to Ho-doped laser crystals. While a detailed analysis of such systems is outside the scope of this Letter, we show in Fig. 5b the absorption spectra of Tm- and Ho-doped YLF crystals and indicate suitable pumping and laser wavelengths. Furthermore, such a two-ion system offers the flexibility of accessing relatively high absorption/gain cross-section transitions for the pump and the laser- as illustrated in Fig. 5b.

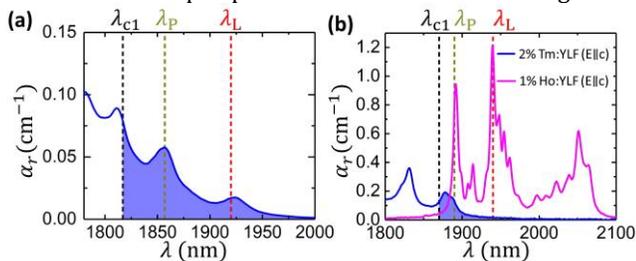

Fig. 5. Proposed pumping schemes for mid-IR (a) RBL in 1% Tm:BYF, and (b) 2T-RBL in a tandem of 2% Tm:YLF and 1% Ho:YLF crystals.

In summary, we report the first demonstration of optical refrigeration in Tm:YLF crystals and present a complete characterization of laser cooling properties of Tm:YLF and Tm:BYF crystals. A 1% Tm:BYF crystal with extremely high EQE ($\eta_{ext}$>99%) and a 2% Tm:YLF crystal with improved resonant absorption show great potential for mid-IR optical cryocooler and RBL applications. The fortunate coincidence between the optimum laser cooling wavelength of Tm:YLF crystals and the efficient pumping wavelength of Ho:YLF laser crystals offers the possibility of realizing high-power mid-IR 2-tone RBLs in the eye-safe region (>1400 nm) of the spectrum.

**Funding.** Air Force Office of Scientific Research (FA9550-15-1-0241 and FA9550-16-1-0362 (MURI)) and U.S. Army "Efficient Materials for Optical Cryocoolers" (W911SR-17-C-0039).
**Acknowledgement.** We thank Drs. Arlete Cassanho and Hans P. Jenssen (AC Materials Inc., Tarpon Springs, FL) for growing high-purity Tm:YLF crystals, and Dr. Richard Epstein (ThermoDynamic Films LLC) and Dr. Zhou Yang (UNM) for insightful discussions.

**References**

1. R. I. Epstein, M. I. Buchwald, B. C. Edwards, T. R. Gosnell, and C. E. Mungan, "Observation of laser-induced fluorescent cooling of a solid," Nature **377**, 500–503 (1995).
2. D. V. Seletskiy, S. D. Melgaard, S. Bigotta, A. Di Lieto, M. Tonelli, and M. Sheik-Bahae, "Laser cooling of solids to cryogenic temperatures," Nat. Photonics **4**, 161–164 (2010).
3. S. Rostami, A. R. Albercht, M. R. Ghasemkhani, S. D. Melgaard, A. Gragossian, M. Tonelli, and M. Sheik-Bahae, "Optical refrigeration of Tm:YLF and Ho:YLF crystals," in *Proc. SPIE 9765, Optical and Electronic Cooling of Solids* (2016), Vol. 9765, p. 97650P–9765–6.
4. S. D. Melgaard, A. R. Albrecht, M. P. Hehlen, and M. Sheik-Bahae, "Solid-state optical refrigeration to sub-100 Kelvin regime," Sci. Rep. **6**, 20380 (2016).
5. M. P. Hehlen, J. Meng, A. R. Albrecht, E. R. Lee, A. Gragossian, S. P. Love, C. E. Hamilton, R. I. Epstein, and M. Sheik-Bahae, "First demonstration of an all-solid-state optical cryocooler," Light Sci. Appl. **7**, 15 (2018).
6. S. Rostami, A. R. Albrecht, A. Volpi, and M. Sheik-Bahae, "Observation of Optical Refrigeration in a Holmium-doped Crystal," arXiv:1811.05745 [physics.optics] (2018).
7. A. Gragossian, A. Volpi, J. Meng, A. R. Albrecht, S. Rostami, M. P. Hehlen, and M. Sheik-Bahae, "Investigation of temperature dependence of quantum efficiency and parasitic absorption in rare-earth doped crystals (Conference Presentation)," in *Proc. SPIE 10550, Optical and Electronic Cooling of Solids III* (2018), Vol. 10550, p. 1055006.
8. C. W. Hoyt, M. Sheik-Bahae, R. I. Epstein, B. C. Edwards, and J. E. Anderson, "Observation of Anti-Stokes Fluorescence Cooling in Thulium-Doped Glass," Phys. Rev. Lett. **85**, 3600–3603 (2000).
9. W. Patterson, S. Bigotta, M. Sheik-Bahae, D. Parisi, M. Tonelli, and R. Epstein, "Anti-Stokes luminescence cooling of Tm^3+ doped BaY_2F_8.," Opt. Express **16**, 1704–1710 (2008).
10. B. Henderson and R. H. Bartram, *Crystal-Field Engineering of Solid-State Laser Materials* (Cambridge University Press, 2000).
11. S. A. Payne, L. L. Chase, L. K. Smith, W. L. Kway, and W. F. Krupke, "Infrared cross-section measurements for crystals doped with Er^3+, Tm^ 3+, and Ho^3+," IEEE J. Quantum Electron. **28**, 2619–2630 (1992).
12. H. P. Jenssen and A. Cassanho, "Fluoride laser crystals: old and new," in *Proc. SPIE 6100, Solid State Lasers XV: Technology and Devices,* (2006), Vol. 6100, p. 61000W.
13. S. D. Melgaard, "Cryogenic optical refrigeration: Laser cooling of solids below 123 K," Ph.D. thesis, University of New Mexico (2013).
14. D. E. McCumber, "Einstein Relations Connecting Broadband Emission and Absorption Spectra," Phys. Rev.

**136**, A954–A957 (1964).
15. M. P. Hehlen, M. Sheik-Bahae, and R. I. Epstein, "Solid-state optical refrigeration," in *Handbook on the Physics and Chemistry of Rare Earths* (2014), Vol. 45, pp. 179–260.
16. S. Bigotta, "Energy transfer and cooling processes in rare-earth doped insulating crystals," Ph.D. thesis, Universit`a degli Studi di Pisa (2004).
17. T. Kessler, C. Hagemann, C. Grebing, T. Legero, U. Sterr, F. Riehle, M. J. Martin, L. Chen, and J. Ye, "A sub-40-mHz-linewidth laser based on a silicon single-crystal optical cavity," Nat. Photonics **6**, 687–692 (2012).
18. O. Graydon, "Payload success," Nat. Photonics **12**, 315 (2018).
19. D. J. Richardson, J. Nilsson, and W. A. Clarkson, "High power fiber lasers: current status and future perspectives [Invited]," J. Opt. Soc. Am. B **27**, B63–B92 (2010).
20. S. R. Bowman, "Lasers without internal heat generation," IEEE J. Quantum Electron. **35**, 115–122 (1999).
21. S. R. Bowman, S. P. O'Connor, S. Biswal, N. J. Condon, and A. Rosenberg, "Minimizing heat generation in solid-state lasers," IEEE J. Quantum Electron. **46**, 1076–1085 (2010).
22. S. R. Bowman, S. O. Connor, S. Biswal, and N. J. Condon, "Demonstration and Analysis of a High Power Radiation Balanced Laser," CLEO 2011 - Laser Sci. to Photonic Appl. 1–2 (2011).
23. Z. Yang, J. Meng, A. R. Albrecht, and M. Sheik-Bahae, "Radiation-balanced Yb:YAG disk laser," Opt. Express **27**, 1392–1400 (2019).
24. S. Rostami, Z. Yang, A. R. Albrecht, A. Gragossian, M. Peysokhan, M. Ghasemkhani, A. Volpi, M. Tonelli, and M. Sheik-Bahae, "Advances in mid-IR solid-state optical cooling and radiation-balanced lasers (Conference Presentation)," in *Proc. SPIE 10550, Optical and Electronic Cooling of Solids III* (2018), Vol. 10550, p. 105500Q.
25. G. Nemova and R. Kashyap, "Radiation-balanced amplifier with two pumps and a single system of ions," J. Opt. Soc. Am. B **28**, 2191–2194 (2011).